\documentclass[twocolumn, secnumarabic, amssymb, nobibnotes, aps, pra]{revtex4}
\usepackage{graphicx}

\usepackage{epsfig,psfrag}

\begin{document}

\title{Solitary-wave description of condensate micro-motion in a \\ time-averaged orbiting potential trap}

\author{K J Challis}
\email{kchallis@physics.otago.ac.nz}
\affiliation{Department of Physics, University of Otago, PO Box 56, Dunedin, New Zealand}
\author{R J Ballagh}
\affiliation{Department of Physics, University of Otago, PO Box 56, Dunedin, New Zealand}
\author{C W Gardiner}
\affiliation{School of Chemical and Physical Sciences, Victoria University, Wellington, New Zealand}

\date{\today}

\begin{abstract}
We present a detailed theoretical analysis of micro-motion in a time-averaged orbiting potential trap.  Our treatment is based on the Gross-Pitaevskii equation, with the full time dependent behaviour of the trap systematically approximated to reduce the trapping potential to its dominant terms.  We show that within some well specified approximations, the dynamic trap has solitary-wave solutions, and we identify a moving frame of reference which provides the most natural description of the system.  In that frame eigenstates of the time-averaged orbiting potential trap can be found, all of which must be solitary-wave solutions with identical, circular centre of mass motion in the lab frame.  The validity regime for our treatment is carefully defined,
and is shown to be satisfied by existing experimental systems.
\end{abstract}

\maketitle

\section{Introduction}

The Time-averaged Orbiting Potential (TOP) trap \cite{Cornell95} was an important tool in the realization of Bose-Einstein condensates, and it remains a common method for magnetically trapping atoms.  Early theoretical descriptions of the TOP trap used two approximations: the adiabatic approximation, which assumes that the magnetic dipoles of the atoms align instantaneously to the magnetic field, and the {\it time-average approximation}, where the time dynamics of the trapping fields are neglected on the time scale of the motion of the trapped atoms.  Under these assumptions, the TOP trap is represented by a static, harmonic potential and the condensate eigenstates are relatively easily calculated (usually by numerical means) and are stationary in space.  However, condensates formed in a TOP trap undergo a spatial micro-motion \cite{Arimondo00,Arimondo01} due to the underlying dynamic nature of the TOP trap.  This phenomenon has been studied theoretically under various levels of approximation, by partially lifting the time-average approximation \cite{Kuklov97,Minogin98} or by not applying the adiabatic approximation \cite{Shtrikman99,Arimondo04}.

In this work we provide a detailed theoretical description of condensate micro-motion in terms of TOP trap eigenstates, including condensate nonlinearity.  Our approach applies the adiabatic approximation, but partially lifts the time-average approximation.  Under these conditions, the TOP trap potential retains some time dependence and eigenstates of that potential cannot be found in the lab frame.  However, system eigenstates do exist because a frame can be found in which the Hamiltonian for the system becomes time independent.  We have termed eigenstates of the system found in such a frame {\it dynamical eigenstates}, since these states are not stationary states in the lab frame.  By calculating the dynamical eigenstates of the TOP trap, full characterisation of condensate micro-motion is possible.  This is essential for an understanding of condensate growth and is also required for a description of velocity sensitive phenomena occuring in TOP traps, such as observed in Bragg scattering experiments \cite{Arimondo02, Wilson03}.  

In this paper we calculate dynamical eigenstates of the TOP trap potential in the {\it quadratic average approximation}.  Within that approximation, the solutions are exact in both the linear {\it and} the nonlinear case.  We begin, in section~\ref{sec:pot}, by introducing the TOP trap potential, and various approximate forms of that potential.  In section~\ref{sec:circ_trans_frame} we derive the transformation to the {\it circularly translating frame} which we find to be the most natural frame in which to investigate the system.  In section~\ref{sec:solitary} we calculate solitary-wave solutions in the quadratic average approximation and show that dynamical eigenstates calculated using the circularly translating frame are a particular class of solitary-wave solutions in the lab frame.  Identifying the dynamical eigenstates of the TOP trap allows us to characterize micro-motion and specify the ground state of the system.  In section~\ref{sec:pert}, we assess the validity of the quadratic average approximation and demonstrate that for the typical parameter regime of the TOP trap, solitary-wave dynamical eigenstates provide accurate approximations to the dynamical eigenstates of the full TOP trap potential.  In section~\ref{sec:rotating}, we discuss lab frame solitary-wave solutions which are eigenstates of the TOP trap potential in the more commonly used rotating frame, and show that these are only a subset of the dynamical eigenstates found using the circularly translating frame.  We conclude in section~\ref{sec:dis}.

\section{Approximate forms of the TOP trap potential \label{sec:pot}}

The TOP trap consists of a magnetic quadrupole trap \cite{Metcalf85,Metcalf87} translated by a uniform bias field, whose direction rotates at frequency $\Omega$ \cite{Cornell95}.  For simplicity our discussion is presented in a set of dimensionless units defined by the position scale $x_0=\sqrt{\hbar/2 m \omega_x}$ (a characteristic harmonic oscillator length), and the time scale of the inverse of the time-averaged trap frequency $\omega_x,$ defined below.  A key feature of the TOP trap is that the zero of the magnetic field follows a circular trajectory of radius $r_0$, and trapped atoms are confined well within this trajectory (the so-called `circle of death'), thereby reducing atom loss due to spin flips.  Typically $r_0 \sim 1000-1300$ and $\Omega \sim 70-150$ \cite{Cornell95,Arimondo00,Wilson03}. 

\subsection{The adiabatic approximation}

The TOP trap potential in the adiabatic approximation is given by
\begin{eqnarray}
V_{\rm{TOP}} (\mbox{\boldmath$r$},t) & = & r_0^2 \left[1+\frac{2 (x \cos\Omega t+y \sin\Omega t)}{r_0} \right. \nonumber \\
& &  \left. +\frac{x^2+y^2+4 z^2}{r_0^2} \right]^{\frac{1}{2}},
\label{TOP_full}
\end{eqnarray}
where $\mbox{\boldmath$r$}=(x,y,z)$.  That approximation is valid when the bias field rotation frequency, $\Omega$, is much smaller than the Lamor precession frequency \cite{Cornell95}.

\subsection{The truncated TOP trap potential}

Expanding the square root of equation (\ref{TOP_full}) in a Taylor series, and neglecting terms above second order in the small parameter $x_{\alpha}/r_0$, where $x_{\alpha}$ is one of $x,$ $y,$ or $z,$ leads to the truncated TOP trap potential
\begin{eqnarray}
V(\mbox{\boldmath$r$},t) & = & r_{0}^2+r_0(x\cos\Omega t+y\sin\Omega t)+\frac{1}{2}(x^2+y^2+4z^2) \nonumber \\
& & -\frac{1}{2}(x\cos \Omega t+y\sin \Omega t)^2.
\label{pot_exp_trunc-3d}
\end{eqnarray}

The evolution of the condensate wave function, $\psi(\mbox{\boldmath$r$},t)$, is governed by the Gross-Pitaevskii equation, 
\begin{equation}
\label{Gross-Pitaevskii equation}
i \frac{\partial}{\partial t} \psi (\mbox{\boldmath$r$},t)=\mathcal{L} (\mbox{\boldmath$r$},t) \psi (\mbox{\boldmath$r$},t),
\end{equation}
where for a TOP trap, the time evolution operator,
\begin{equation}
\label{Lop}
\mathcal{L} (\mbox{\boldmath$r$},t)=-\nabla ^{2}+V(\mbox{\boldmath$r$},t)+C|\psi(\mbox{\boldmath$r$},t)|^2,
\end{equation}
is time dependent.  The truncated TOP trap potential $V(\mbox{\boldmath$r$},t)$ is given by equation (\ref{pot_exp_trunc-3d}), and $C$ is the dimensionless nonlinearity coefficient defined in terms of the number of atoms $N$, and the s-wave scattering length $a$, i.e.
\begin{equation}
C=\frac{4\pi\hbar a N}{m \omega_x x_0^3}.
\label{Cvalue}
\end{equation}

\subsection{The time-average approximation}

The most common treatment of condensate evolution in a TOP trap has also invoked the time-average approximation, whereby the potential of equation (\ref{pot_exp_trunc-3d}) is averaged over a period of the bias field rotation.  This leads to the time-averaged, truncated form of the TOP trap potential, 
\begin{equation}
\label{harm_pot}
V_{\rm{H}}(\mbox{\boldmath$r$})=r_0^2+\frac{1}{4}(x^2+y^2+8z^2),
\end{equation}
which is a static, harmonic potential, with frequency $\omega_x$ in the $x$-$y$ plane (in SI units).  In the time-average approximation, the potential in the time evolution operator of equation (\ref{Lop}), is replaced by the time independent trap $V_{\rm{H}}(\mbox{\boldmath$r$}),$ of equation (\ref{harm_pot}).  This allows energy eigenstates of the system to be readily calculated.  

The time-average approximation is normally assumed to be valid when the bias field rotation frequency is much larger than the frequency of the time-averaged harmonic trap, i.e.\ in our dimensionless units $\Omega \gg 1.$  The time-averaged treatment neglects system dynamics occurring on the fast time scale of the bias field rotation, and it is this non-stationary behaviour of a condensate in a TOP trap that we describe in this work.

\subsection{The quadratic average approximation}

M$\rm{\ddot{u}}$ller et al.\ \cite{Arimondo00} experimentally observed the dynamic effects of the TOP trap on condensate evolution, i.e.\ micro-motion in a TOP trap.  Their approach for calculating the condensate micro-motion amplitude involved balancing the restoring force of the time dependent terms of equation (\ref{pot_exp_trunc-3d}) that are linear in $\cos \Omega t$ or $\sin \Omega t,$ with the centrifugal force.  In line with this treatment, our work invokes what we shall refer to as the {\it quadratic average approximation}, where only the terms of equation (\ref{pot_exp_trunc-3d}) that are quadratic in $\cos \Omega t$ or $\sin \Omega t$ are time-averaged.  In that approximation the TOP trap potential is given by
\begin{eqnarray}
\label{approx-pot-3d}
V_{\rm{ap}}(\mbox{\boldmath$r$},t) & = & V_{\rm{H}}(\mbox{\boldmath$r$})+r_0(x\cos\Omega t+y\sin\Omega t).
\end{eqnarray}

In the present paper, we calculate dynamical eigenstates of the TOP trap potential in the quadratic average approximation, where the trapping potential is given by equation (\ref{approx-pot-3d}).  The accuracy of the quadratic average approximation is addressed in section~\ref{sec:pert}.

\section{The circularly translating frame \label{sec:circ_trans_frame}}

In the lab frame, the TOP trap potential in the quadratic average approximation, given by equation (\ref{approx-pot-3d}), is time dependent and eigenstates of the Gross-Pitaevskii equation can not be found.   By transforming to a frame that translates in a circular trajectory with radius $\gamma$ (whose value is to be determined) and with angular frequency $\Omega$ about the origin of the lab frame, we can remove this time dependence.  We refer to that frame as the {\it circularly translating frame} and we shall see that it is the natural frame in which to describe the TOP trap system.

\begin{figure}
\includegraphics[width=1.83in]{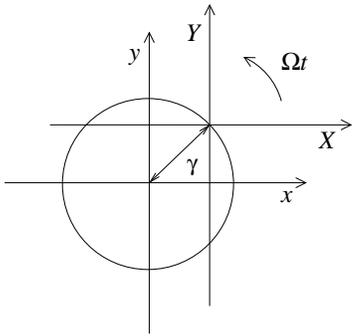}
\caption{\label{fig:circ_trans}  The circularly translating frame, defined by co-ordinates $\mbox{\boldmath$R$}=(X,Y,Z)$ and equation (\ref{circtrans}). }
\end{figure}

The translation in co-ordinate space is defined by 
\begin{equation}
\mbox{\boldmath$R$} = \mbox{\boldmath$r$} -\gamma (\cos \Omega t, \sin \Omega t,0), \label{circtrans}
\end{equation}
as illustrated in figure \ref{fig:circ_trans}.
The momentum in the circularly translating frame is derived by differentiating equation (\ref{circtrans}), yielding
\begin{equation}
\mbox{\boldmath$P$}  = \mbox{\boldmath$p$} +\frac{1}{2}\gamma \Omega (\sin \Omega t, -\cos \Omega t,0),
\label{circtrans_p}
\end{equation}
where we have used the fact that in our dimensionless units $\mbox{\boldmath$p$}=\mbox{\boldmath$v$}/2.$

\subsection{Unitary transformation to the circularly translating frame}

We now derive the quantum mechanical transformation to the circularly translating frame.  For clarity, we shall denote quantum mechanical operators by $\hat{O},$ and begin with the linear case. 

The Schr$\rm{\ddot{o}}$dinger equation for a single particle state in the TOP trap, in the quadratic average approximation, is given in the lab frame by \begin{equation}
i \frac{d|\psi \rangle}{dt}=\hat{H}_{\rm{ap}} |\psi \rangle,
\label{Schrodinger_eqn}
\end{equation}
where
\begin{equation}
\hat{H}_{\rm{ap}}=\hat{\mbox{\boldmath$p$}}^2+\hat{V}_{\rm{ap}} (\hat{\mbox{\boldmath$r$}},t).
\end{equation}
The transformation to the circularly translating frame is achieved by the unitary transformation 
\begin{equation}
\hat{U}(t) \equiv \hat{U}_{\mbox{\boldmath$p$}} (\mbox{\boldmath$b$}(t)) \hat{U}_{\mbox{\boldmath$r$}} (\mbox{\boldmath$a$}(t)),
\end{equation}
where 
\begin{equation}
\hat{U}_{\mbox{\boldmath$r$}} (\mbox{\boldmath$a$}(t))=e ^{i \hat{\mbox{\boldmath$p$}} \cdot \mbox{\boldmath$a$}(t)}
\end{equation}
translates position by $\mbox{\boldmath$a$}(t),$ $\hat{U}_{\mbox{\boldmath$r$}} (\mbox{\boldmath$a$}(t))|\mbox{\boldmath$r$}\rangle=|\mbox{\boldmath$r$}-\mbox{\boldmath$a$}(t)\rangle,$
and
\begin{equation}
\hat{U}_{\mbox{\boldmath$p$}} (\mbox{\boldmath$b$}(t))=e ^{-i \hat{\mbox{\boldmath$r$}} \cdot \mbox{\boldmath$b$}(t)},
\end{equation}
translates momentum by $\mbox{\boldmath$b$}(t),$ $\hat{U}_{\mbox{\boldmath$p$}} (\mbox{\boldmath$b$}(t))|\mbox{\boldmath$p$}\rangle=|\mbox{\boldmath$p$}-\mbox{\boldmath$b$}(t)\rangle.$  In a comparison with equations (\ref{circtrans}) and (\ref{circtrans_p}) we find that
\begin{equation}
\mbox{\boldmath$a$}(t)=\gamma (\cos\Omega t,\sin\Omega t,0),
\label{aeqn}
\end{equation}
and
\begin{equation}
\mbox{\boldmath$b$}(t)=-\frac{1}{2}\gamma\Omega (\sin\Omega t,-\cos\Omega t,0).
\label{beqn}
\end{equation}
In the transformation to the circularly translating frame $\hat{U}_{\mbox{\boldmath$r$}} (\mbox{\boldmath$a$}(t))$ and $\hat{U}_{\mbox{\boldmath$p$}} (\mbox{\boldmath$b$}(t))$ commute, since $\mbox{\boldmath$a$}(t) \cdot \mbox{\boldmath$b$}(t)=0$.  Defining the transformed state vector to be
\begin{equation}
|\psi\rangle ^{\rm{t}}\equiv \hat{U}(t) |\psi \rangle,
\label{def_vector}
\end{equation}
equation (\ref{Schrodinger_eqn}) becomes 
\begin{equation}
i \frac{d|\psi \rangle ^{\rm{t}}}{dt}=\hat{H}_{\rm{ap}}^{\rm{t}} |\psi \rangle ^{\rm{t}},
\label{SE-trans}
\end{equation}
where
\begin{eqnarray}
\hat{H}_{\rm{ap}}^{\rm{t}} & = & \hat{U}_{\mbox{\boldmath$p$}}(\mbox{\boldmath$b$}(t)) \hat{U}_{\mbox{\boldmath$r$}}(\mbox{\boldmath$a$}(t)) \hat{H}_{\rm{ap}} \hat{U}^{\dagger}_{\mbox{\boldmath$r$}}(\mbox{\boldmath$a$}(t)) \hat{U}^{\dagger}_{\mbox{\boldmath$p$}}(\mbox{\boldmath$b$}(t)) \nonumber \\ & & -\left( \hat{\mbox{\boldmath$p$}}+\mbox{\boldmath$b$}(t) \right) \cdot \frac{d\mbox{\boldmath$a$}(t)}{dt}+\hat{\mbox{\boldmath$r$}}\cdot \frac{d \mbox{\boldmath$b$}(t)}{dt}.
\end{eqnarray}
The Schr$\rm{\ddot{o}}$dinger equation in the co-ordinate representation can be determined by projecting equation (\ref{SE-trans}) onto state $|\mbox{\boldmath$R$} \rangle.$  Using the following identities,
\begin{equation}
\langle \mbox{\boldmath$R$} |\psi \rangle = \psi (\mbox{\boldmath$R$},t),
\end{equation}
and
\begin{equation}
\langle \mbox{\boldmath$R$}|\hat{\mbox{\boldmath$p$}} |\psi \rangle = -i \nabla _{\mbox{\boldmath$R$}} \psi (\mbox{\boldmath$R$},t),
\end{equation} 
where we have denoted
\begin{equation}
\nabla _{\mbox{\boldmath$R$}}=\left( \frac{\partial}{\partial X},\frac{\partial}{\partial Y},\frac{\partial }{\partial Z} \right),
\end{equation}
yields
\begin{equation}
i \frac{\partial \psi ^{\rm{t}}(\mbox{\boldmath$R$},t)}{\partial t} =H_{\rm{ap}}^{\rm{t}}(\mbox{\boldmath$R$},t) \psi ^{\rm{t}}(\mbox{\boldmath$R$},t),
\end{equation}
where
\begin{eqnarray}
H_{\rm{ap}}^{\rm{t}}(\mbox{\boldmath$R$},t) & = & \left( -i \nabla _{\mbox{\boldmath$R$}}+\mbox{\boldmath$b$}(t) \right) ^2 +V_{\rm{ap}} (\mbox{\boldmath$R$}+\mbox{\boldmath$a$}(t)) \nonumber \\
& & +\left(i \nabla _{\mbox{\boldmath$R$}}-\mbox{\boldmath$b$}(t)\right) \cdot \frac{d\mbox{\boldmath$a$}(t)}{dt} \nonumber \\
& & + \mbox{\boldmath$R$} \cdot \frac{d\mbox{\boldmath$b$}(t)}{dt}.
\label{Hamiltonian_full}
\end{eqnarray}
The wave functions in the lab frame and the circularly translating frame are related by 
\begin{equation}
\psi ^{\rm{t}} (\mbox{\boldmath$R$},t)=e ^{-i \mbox{\boldmath$R$}\cdot \mbox{\boldmath$b$} (t)}\psi (\mbox{\boldmath$R$}+\mbox{\boldmath$a$} (t),t).
\label{int_picture}
\end{equation}

\subsection{Application to the Gross-Pitaevskii equation}

The above derivation, for the Schr$\rm{\ddot{o}}$dinger equation, may also be adapted to the Gross-Pitaevskii equation, since the nonlinear term transforms simply under substitution of equation (\ref{int_picture}).  Thus, by substituting $\mbox{\boldmath$a$}(t)$ and $\mbox{\boldmath$b$}(t)$ from equations (\ref{aeqn}) and (\ref{beqn}) into the Hamiltonian of equation (\ref{Hamiltonian_full}), and including the nonlinear term (which is described in terms of the new density $|\psi ^{\rm{t}} (\mbox{\boldmath$R$},t)|^2$), the Gross-Pitaevskii equation in the circularly translating frame is
\begin{equation}
\label{Gross-Pitaevskii equation-ct}
i \frac{\partial}{\partial t} \psi^{\rm{t}} (\mbox{\boldmath$R$},t)=\mathcal{L}_{\rm{ap}}^{\rm{t}} (\mbox{\boldmath$R$},t) \psi^{\rm{t}} (\mbox{\boldmath$R$},t),
\end{equation}
where 
\begin{equation}
\mathcal{L}_{\rm{ap}}^{\rm{t}} (\mbox{\boldmath$R$},t)  =  H_{\rm{ap}}^{\rm{t}}(\mbox{\boldmath$R$},t)
+C|\psi^{\rm{t}}(\mbox{\boldmath$R$},t)|^2,
\label{Lop-ct-gen}
\end{equation}
and
\begin{eqnarray}
H_{\rm{ap}}^{\rm{t}}(\mbox{\boldmath$R$},t) & = &  - \nabla _{\mbox{\boldmath$R$}} ^2+V_{\rm{H}}(\mbox{\boldmath$R$}) \nonumber \\
& & +\frac{1}{2}(\gamma+2 r_0-\gamma\Omega^2)(X\cos\Omega t+Y \sin\Omega t) \nonumber \\ & & +\gamma r_0 -\frac{1}{4}\gamma^2 (\Omega^2-1).
\label{H-ct-gen}
\end{eqnarray}
The single particle Hamiltonian of equation (\ref{H-ct-gen}) is identical to the Hamiltonian derived using a classical frame transformation to a non-inertial frame of reference \cite{L&L}, applied to the circularly translating frame.   Choosing $\gamma=\gamma_{\rm{t}},$ where 
\begin{equation}
\gamma_{\rm{t}}=\frac{2r_0}{\Omega^2-1},
\label{gamma}
\end{equation}
the evolution operator of equation (\ref{Lop-ct-gen}) simplifies to 
\begin{equation}
\label{Lop-ct}
\mathcal{L}_{\rm{ap}}^{\rm{t}} (\mbox{\boldmath$R$},t)= -\nabla _{\mbox{\boldmath$R$}} ^2 +V_{\rm{H}}(\mbox{\boldmath$R$})+C|\psi^{\rm{t}}(\mbox{\boldmath$R$},t)|^2+\varepsilon,
\end{equation}
where 
\begin{equation}
\varepsilon = \frac{1}{4} \gamma_{\rm{t}} ^2 (\Omega ^2-1).
\label{epsilon-algebra}
\end{equation}
The energy offset $\varepsilon$ can be interpreted by expressing equation (\ref{epsilon-algebra}) in the form
\begin{eqnarray}
\varepsilon & = & V_{\rm{ap}}([\gamma_{\rm{t}},0,0],t=0)-V_{\rm{H}} (\mathbf{0})-E_{\Omega } 
\label{varepsilon}.
\end{eqnarray}
The first two terms represent the additional potential energy due to the displacement of a point body from the trap centre to radius $\gamma_{\rm{t}}.$  The remaining term, $E_{\Omega}=\gamma_{\rm{t}} ^2 \Omega ^2 /4,$ represents the energy of a point body rotating about the origin of the lab frame, at a radius $\gamma_{\rm{t}}$ with frequency $\Omega,$ which is simply the expected energy shift associated with the transformation to the circularly translating frame \cite{L&L}. 

The time evolution operator in the circularly translating frame, as given by equation (\ref{Lop-ct}), substituted into equation (\ref{Gross-Pitaevskii equation-ct}) yields the Gross-Pitaevskii equation for a time independent harmonic trap, with an energy offset of $\varepsilon.$  Thus, eigenstates of the TOP trap in the circularly translating frame exist in the quadratic average approximation.  For clarity we write these as 
\begin{equation}
\psi ^{\rm{t}} (\mbox{\boldmath$R$},t)=\psi _{\rm{H}} (\mbox{\boldmath$R$})e ^{-i (\mu_{\rm{H}}+\varepsilon) t},
\label{estates_ctf}
\end{equation}
where $\psi_{\rm{H}} (\mbox{\boldmath$R$})$ are the well-known solutions to the time independent Gross-Pitaevskii equation for a time independent harmonic trap, i.e.
\begin{equation}
\label{Gross-Pitaevskii equation-psi-H}
\mu_{\rm{H}}\psi_{\rm{H}} (\mbox{\boldmath$R$})=\left[ -\nabla_{\mbox{\boldmath$R$}} ^{2}+V_{\rm{H}}(\mbox{\boldmath$R$})+C|\psi_{\rm{H}}(\mbox{\boldmath$R$})|^2 \right] \psi_{\rm{H}}(\mbox{\boldmath$R$}).
\end{equation}

\subsection{Generalization to quantum field theory \label{sec:qft}}

The transformation given by equation (\ref{int_picture}) can be applied to the operator Heisenberg equations of motion for the full quantum field operator 
$\hat{\Psi}(\mbox{\boldmath$r$},t)$.  In much the same way as our discussion above, this yields the equation of motion in the circularly translating frame
\begin{eqnarray}
i \frac{\partial \hat{\Psi}_{\rm{t}}(\mbox{\boldmath$R$},t)}{\partial t} & = & 
\left[ -\nabla _{\mbox{\boldmath$R$}} ^2 
+V_{\rm{H}}(\mbox{\boldmath$R$})+C\hat{\Psi}^\dagger_{\rm{t}}(\mbox{\boldmath$R$},t) \hat{\Psi}_{\rm{t}}(\mbox{\boldmath$R$},t) \right. \nonumber \\
& & \left. +\varepsilon \right]  \hat{\Psi}_{\rm{t}}(\mbox{\boldmath$R$},t).
\end{eqnarray}
Since this represents the full quantum field theory, the motion of uncondensed particles is also correctly treated in the circularly translating frame.  

\section{Solitary-wave solutions \label{sec:solitary}}

Solitary-wave solutions, where the wave function evolves without changing shape, can be found for the TOP trap in the quadratic average approximation.  Morgan et al.\ \cite{Morgan97} have shown that the Gross-Pitaevskii equation, with particular forms of potential, has solitary-wave solutions which propagate in one dimension of a multi-dimensional space.  That work was extended by Margetis \cite{Margetis99} where solitary-wave solutions may have center of mass motion in any of the space dimensions.  Also, Japha and Band \cite{Band02} have shown that in a moving harmonic trap the motion of the condensate centre of mass can be entirely decoupled from the evolution of the condensate shape.  We have extended the derivation by Morgan et al.\ \cite{Morgan97} to include the case where solitary-wave solutions can propagate in three dimensions, as was indicated to be possible by Margetis \cite{Margetis99}.  In the following we present a brief summary of the results of our derivation.

We begin by postulating that solitary-wave solutions to the TOP trap will have the form 
\begin{equation}
\label{psi-postulate}
\psi_{\rm{SW}}(\mbox{\boldmath$r$},t)=\psi_{\rm{H}}(\mbox{\boldmath$r$}-\bar{\mbox{\boldmath$r$}}(t))e ^{-i \mu _{\rm{H}} t+i S(\mbox{\boldmath $r$},t)},
\end{equation}
where the envelope wave function $\psi_{\rm{H}}(\mbox{\boldmath$r$})$ is an eigenstate of the time independent Gross-Pitaevskii equation for the TOP trap potential in the time-average approximation with chemical potential $\mu_{\rm{H}}$, i.e.\ defined by equation (\ref{Gross-Pitaevskii equation-psi-H}).  The position offset in the envelope wave function is
\begin{eqnarray}
\bar{\mbox{\boldmath$r$}}(t)& = & \int \psi^*_{\rm{SW}}(\mbox{\boldmath$r$},t) \mbox{\boldmath$r$} \psi_{\rm{SW}}(\mbox{\boldmath$r$},t)
d \mbox{\boldmath$r$} \nonumber \\
& & -\int \psi^*_{\rm{H}}(\mbox{\boldmath$r$}) \mbox{\boldmath$r$} \psi_{\rm{H}}(\mbox{\boldmath$r$})
d \mbox{\boldmath$r$},
\end{eqnarray}
which can be interpreted as the time dependent position of the centre of mass of the solitary-wave since the second integral is zero due to the particular form of $V_{\rm{H}}(\mbox{\boldmath$r$}).$  The phase $S(\mbox{\boldmath$r$},t)$ is determined by substituting the solitary-wave solution, equation (\ref{psi-postulate}), into the time dependent Gross-Pitaevskii equation, equation (\ref{Gross-Pitaevskii equation}), where $\mathcal{L} (\mbox{\boldmath$r$},t)$ is replaced by $\mathcal{L}_{\rm{ap}} (\mbox{\boldmath$r$},t),$ in which the quadratic average approximation is used, i.e.
\begin{equation}
\label{Lop_ap}
\mathcal{L}_{\rm{ap}} (\mbox{\boldmath$r$},t)=-\nabla ^{2}+V_{\rm{ap}}(\mbox{\boldmath$r$},t)+C|\psi(\mbox{\boldmath$r$},t)|^2,
\end{equation}
where $V_{\rm{ap}}(\mbox{\boldmath$r$},t)$ is given by equation (\ref{approx-pot-3d}).  Taking a similar approach to that of Morgan et al.\ \cite{Morgan97} the Gross-Pitaevskii equation can be separated into real and imaginary parts yielding two equations.  The equation derived from the imaginary part can be simplified by writing
\begin{equation}
S(\mbox{\boldmath$r$},t)=\frac{1}{2} \mbox{\boldmath$r$} \cdot \frac{d \bar{\mbox{\boldmath$r$}}(t)}{dt}+K(\mbox{\boldmath$r$},t).
\label{Sfunc}
\end{equation}
We choose the trivial solution, $K(\mbox{\boldmath$r$},t) = K(t)$, which is the only possible solution in the one dimensional case \cite{Morgan97} and has also been suggested as the unique solution in the general case \cite{Margetis99}.  Substituting the trivial solution into the equation derived from the real part of the Gross-Pitaevskii equation we find that 
\begin{equation}
K(t)  =  \frac{1}{4}\int  \left[ \bar{\mbox{\boldmath$r$}}^2(t)-\left( \frac{d \bar{\mbox{\boldmath$r$}}(t)}{dt} \right)^2 \right] dt.
\label{Kfunc}
\end{equation}
By equating mixed differentials of $S(\mbox{\boldmath$r$},t)$, the centre of mass motion of the solitary-wave solutions can be found to obey
\begin{equation}
\frac{1}{2}\frac{\partial ^2 \bar{\mbox{\boldmath$r$}}(t)}{\partial t^2}=-\nabla F (\mbox{\boldmath$r$},t),
\label{Ehrenfest}
\end{equation}
which is a form of Ehrenfest's theorem, where 
\begin{equation}
F(\mbox{\boldmath$r$},t)=V_{\rm{ap}}(\mbox{\boldmath$r$},t)-V_{\rm{H}}(\mbox{\boldmath$r$}-\bar{\mbox{\boldmath$r$}}(t)).
\end{equation}
For solitary-wave solutions to exist, in the form that we have discussed, both sides of equation (\ref{Ehrenfest}) must be independent of $\mbox{\boldmath$r$}$ and therefore the function $F(\mbox{\boldmath$r$},t)$ must be at most linear in $\mbox{\boldmath$r$}.$  The TOP trap potential in the quadratic average approximation obeys this criteria and thus solitary-wave solutions exist.  It is possible to solve for $\bar{\mbox{\boldmath$r$}}(t)$ which has six constants of integration, given by the initial values of the centre of mass position and momentum of the particular solitary-wave solution (see equation (\ref{centre_mass_motion})).

\subsection{Solitary-wave solutions which are eigenstates in the circularly translating frame}

The dynamical eigenstates calculated using the circularly translating frame, given by equation (\ref{estates_ctf}), are a particular class of solitary-wave solutions in the lab frame.  This can be confirmed by transforming the solitary-wave solutions, as given by equation (\ref{psi-postulate}), into the circularly translating frame, and requiring that these solutions satisfy the time independent Gross-Pitaevskii equation in that frame, i.e.
\begin{equation}
\mu_{\rm{SW}}^{\rm{t}} \psi _{\rm{SW}}^{\rm{t}} (\mbox{\boldmath$R$},t)=\mathcal{L}_{\rm{ap}}^{\rm{t}} (\mbox{\boldmath$R$},t) \psi _{\rm{SW}} ^{\rm{t}} (\mbox{\boldmath$R$},t),
\label{Gross-Pitaevskii equation_tind}
\end{equation}
where
\begin{eqnarray}
\psi _{\rm{SW}}^{\rm{t}} (\mbox{\boldmath$R$},t) & = & \psi _{\rm{H}} (\mbox{\boldmath$r$}- \bar{\mbox{\boldmath$r$}}(t)) e ^{-i \mu_{\rm{H}}t +i S(\mbox{\boldmath$r$},t)} \nonumber \\
& & e^{\frac{1}{2} i  \gamma_{\rm{t}} \Omega \mbox{\boldmath$r$}\cdot (\sin\Omega t,-\cos\Omega t,0)},
\label{solution_ctframe}
\end{eqnarray}
and 
\begin{equation}
\mu _{\rm{SW}}^{\rm{t}}  =  \mu_{\rm{H}}+\varepsilon.
\label{mu-ct} 
\end{equation}
In order to satisfy equation (\ref{Gross-Pitaevskii equation_tind}), the solitary-wave solutions of equation (\ref{solution_ctframe}) have a restriction on $\bar{\mbox{\boldmath$r$}}(t)$, as derived in appendix \ref{appendix:ctf}.  Solitary-wave solutions of the lab frame are eigenstates of the TOP trap in the circularly translating frame if and only if the initial conditions of the centre of mass motion of the solitary-wave solutions have particular values such that
\begin{equation}
\label{meanx_col}
\bar{\mbox{\boldmath$r$}}(t)=\gamma_{\rm{t}} (\cos \Omega t, \sin \Omega t ,0),
\end{equation}
and therefore
\begin{equation}
\label{meank_col}
\bar{\mbox{\boldmath$p$}}(t) =-\frac{1}{2}\gamma_{\rm{t}}\Omega (\sin \Omega t, -\cos \Omega t, 0).
\end{equation}
These equations represent circular motion at the TOP trap frequency $\Omega  $, with radius $ \gamma_{\rm{t}}$.  Previously \cite{Arimondo00,Wilson03} the micro-motion position amplitude has been determined to be $2r_0/\Omega^2$, which is in agreement with our result (see equation (\ref{gamma})) in the limit $\Omega \gg 1$.  

With the centre of mass motion for solitary-wave dynamical eigenstates of the TOP trap given by equations (\ref{meanx_col}) and (\ref{meank_col}), we find that $\mbox{\boldmath$R$}=\mbox{\boldmath$r$}-\bar{\mbox{\boldmath$r$}}(t)$ and the phase $S(\mbox{\boldmath$r$},t)$ becomes
\begin{equation}
S(\mbox{\boldmath$r$},t)=-\frac{1}{2} \gamma_{\rm{t}}\Omega \mbox{\boldmath$r$} \cdot (\sin\Omega t,-\cos\Omega t,0) -\varepsilon t.
\end{equation}
Making these substitutions, equation (\ref{solution_ctframe}) simplifies to equation (\ref{estates_ctf}) so that all dynamical eigenstates of the TOP trap, calculated using the circularly translating frame, are a particular class of solitary-wave solutions in the lab frame with centre of mass motion given by equations (\ref{meanx_col}) and (\ref{meank_col}).  This shows that the origin of the circularly translating frame (refer to equations (\ref{circtrans}) and (\ref{circtrans_p}) with $\gamma=\gamma_{\rm{t}}$) moves with the centre of mass motion of the solitary-wave dynamical eigenstates of the TOP trap, therefore justifying our choice of the circularly translating frame for describing the TOP trap.  

\subsection{Dynamical eigenstates \label{sec:eigenstates}}

All the dynamical eigenstates of the TOP trap follow the same circular trajectory in the lab frame, as given by equations (\ref{meanx_col}) and (\ref{meank_col}).  This motion is independent of both the chemical potential of the state and the nonlinear strength of the system.  Furthermore, the solitary-wave dynamical eigenstates retain their orientation with the lab frame throughout their trajectory.

\begin{figure}
\includegraphics[width=3in]{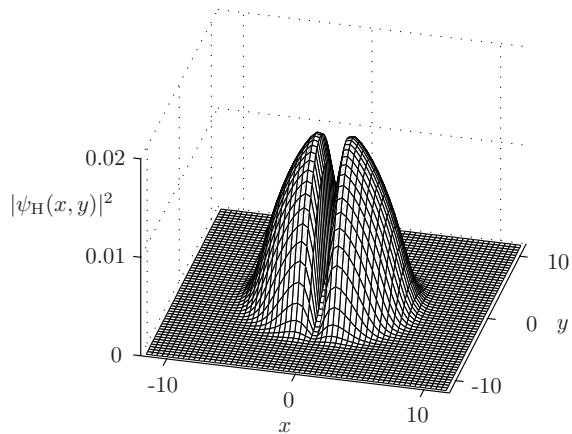}
\caption{\label{fig:exstate}  A particular eigenstate of the two dimensional equivalent of equation (\ref{Gross-Pitaevskii equation-psi-H}), calculated numerically.  The two dimensional nonlinear strength is $C_{\rm 2D}=600$ and the chemical potential is $\mu_{\rm{H}}=10.56$. }
\end{figure}

As an example, a two dimensional excited state of the envelope wave function $\psi_{\rm{H}}(x,y)$, with a nodal line along the $y$ axis, is presented in figure \ref{fig:exstate}. 
In the lab frame, the solitary-wave dynamical eigenstate of the TOP trap, corresponding to the envelope wave function in figure (\ref{fig:exstate}), consists of the envelope wave function moving in a circular trajectory, while maintaining its orientation with the lab frame, and the orientation of the nodal line along the $y$ axis.

In the linear case ($C=0$) the solitary-wave dynamical eigenstates form a complete basis for the TOP trap in the circularly translating frame.  This is because these states, in the circularly translating frame, are eigenstates of the harmonic oscillator equation with an additional energy offset (see equations (\ref{Lop-ct}) and (\ref{estates_ctf})). 

\subsection{Condensation and the ground state of the TOP trap}

The ground state of the TOP trap system is the solitary-wave dynamical eigenstate with the lowest chemical potential in the circularly translating frame.  This occurs when $\mu_{\rm{H}},$ the chemical potential of the envelope wave function, takes its lowest possible value (see equation (\ref{mu-ct})).  Since, as noted in section~\ref{sec:qft}, the uncondensed atoms experience the same potential as the condensate, in the circularly translating frame, these thermalize during evaporation into the usual Bose-Einstein distribution, and hence condensation from the vapour will be into the TOP trap ground state, as determined using the circularly translating frame.  Therefore, the state into which bosons condense, in the quadratic average approximation is the solitary-wave dynamical eigenstate given by equation (\ref{solution_ctframe}) with the envelope wave function being the ground state of equation (\ref{Gross-Pitaevskii equation-psi-H}) and $\bar{\mbox{\boldmath$r$}}(t)$ given by equation (\ref{meanx_col}).

\section{Validity of approximations\label{sec:pert}}

\subsection{Corrections to the quadratic average approximation}

Throughout this work we have used the quadratic average approximation without assessing its validity.  Here we give a systematic assessment of the validity regime of the quadratic average approximation for the linear case (where the mean-field interaction is neglected).  This allows simple analytic results to be obtained.  

The single particle Hamiltonian, with the truncated time dependent TOP trap potential of equation (\ref{pot_exp_trunc-3d}), takes the form (in the circularly translating frame) 
\begin{equation}
\label{hamil-ct-op}
\hat{H}^{\rm{t}}= \hat{H}^{\rm{t}}_{\rm{ap}}+\hat{W}(\hat{\mbox{\boldmath$R$}}),
\end{equation}
where $\hat{H}^{\rm{t}}_{\rm{ap}}$ is the single particle Hamiltonian in the quadratic average approximation, i.e.
\begin{equation}
\hat{H}^{\rm{t}}_{\rm{ap}}=\hat{P}_{X}^2+\hat{P}_{Y}^2+\hat{P}_{Z}^2+\hat{V}_{\rm{H}} (\hat{\mbox{\boldmath$R$}})+\varepsilon.
\label{H-qaa}
\end{equation}
In equations (\ref{hamil-ct-op}) and (\ref{H-qaa}) we have used an operator formalism where the position and momentum component operators are denoted by $\hat{\mbox{\boldmath$R$}}$ and  ${\hat{P}}_{\rm{J}},$ respectively.  The perturbative potential, $\hat{W}(\hat{\mbox{\boldmath$R$}})$, accounts for the remaining terms of the TOP trap potential of equation (\ref{pot_exp_trunc-3d}) that are not retained in the quadratic average approximation.  In the circularly translated frame, these terms are given by
\begin{eqnarray}
\hat{W}(\hat{\mbox{\boldmath$R$}}) & = & -\frac{1}{4}(\hat{X}^2-\hat{Y}^2) (\cos ^2 \Omega t-\sin ^2 \Omega t) \nonumber \\ & & -\frac{1}{2}\gamma_{\rm{t}} (\hat{X} \cos\Omega t+\hat{Y} \sin\Omega t) \nonumber \\ & & -\hat{X} \hat{Y} \sin\Omega t \cos\Omega t-\frac{1}{4}\gamma_{\rm{t}} ^2.
\label{pert_pot}
\end{eqnarray}

The harmonic oscillator creation and annihilation operators in the circularly translating frame, are defined in our dimensionless units, as
\begin{eqnarray}
a_{X} & = & \frac{1}{2}\hat{X}+i \hat{P}_{X} \\
a_{Y} &  = & \frac{1}{2}\hat{Y}+i \hat{P}_{Y} \\
a_{Z} &  = & \sqrt{2}\hat{Z}+i \hat{P}_{Z},
\end{eqnarray}
where $[a_{J},a_{K}^{\dagger}]=\delta_{JK}$, and $J$ and $K$ are one of $X$, $Y,$ or $Z$.  Making these substitutions we find that
\begin{equation}
\hat{H}^{\rm{t}}_{\rm{ap}}=a_X^{\dagger}a_X+a_Y^{\dagger}a_Y+a_Z^{\dagger}a_Z+\varepsilon+r_{\rm{0}}^2+1+\sqrt{2},
\end{equation}
and
\begin{eqnarray} 
\hat{W}(\hat{\mbox{\boldmath$R$}}) & = &  -\frac{1}{4}(a_X^{\dagger 2}+a_X^2+2a_X^{\dagger}a_X) (\cos ^2 \Omega t-\sin ^2 \Omega t) 
\nonumber \\  & &  +\frac{1}{4}(a_Y^{\dagger 2}+a_Y^2+2a_Y^{\dagger}a_Y) (\cos ^2 \Omega t-\sin ^2 \Omega t) 
\nonumber \\  & & 
-\frac{1}{2}\gamma_{\rm{t}} (a_X^{\dagger}+a_X) \cos\Omega t-
 \frac{1}{2}\gamma_{\rm{t}} (a_Y^{\dagger}+a_Y) \sin\Omega t 
\nonumber \\ & & 
-(a_X^{\dagger}+a_X)(a_Y^{\dagger}+a_Y) \sin\Omega t \cos\Omega t \nonumber \\ & & -\frac{1}{4}\gamma_{\rm{t}} ^2.
\label{pert_pot_op}
\end{eqnarray}

Utilising the number operator kets, which satisfy $a_J^{\dagger}a_J |n_J \rangle = n_J |n_J \rangle,$ the eigenket of the single particle Hamiltonian is $|n_X,n_Y,n_Z \rangle ^{\rm{t}},$ i.e.
\begin{equation}
\hat{H}^{\rm{t}}_{\rm{ap}} |n_X,n_Y,n_Z \rangle ^{\rm{t}} = E^{\rm{t}} |n_X,n_Y,n_Z \rangle ^{\rm{t}}.
\end{equation}
The energy spectrum is given by 
\begin{equation}
E^{\rm{t}}=n_X+n_Y+n_Z+\varepsilon+r_{\rm{0}}^2+1+\sqrt{2},
\end{equation}
in agreement with equation (\ref{mu-ct}).  The energy spectrum of the harmonic oscillator terminates at the ground state, $|0,0,0 \rangle ^{\rm{t}},$ which has energy $E_{\rm{g}}^{\rm{t}}=\varepsilon+r_{\rm{0}}^2+1+\sqrt{2}.$

Using time dependent perturbation theory, the evolution of the ground state to first order in the perturbation $\hat{W}(\hat{\mbox{\boldmath$R$}})$ is given by
\begin{eqnarray}
|\psi \rangle ^{\rm{t}} & = & A(t)e ^{-i (E_{\rm{g}}^{\rm{t}}-\frac{1}{4}\gamma_{\rm{t}}^2)t}
\left[ |0,0,0 \rangle^{\rm{t}} \right. \nonumber \\ 
& & +\frac{\gamma_{\rm{t}}}{4}\left( \frac{2 e ^{-i t}}{\Omega^2-1}+ \frac{e ^{i\Omega t}}{\Omega+1}-\frac{e^{-i \Omega t}}{\Omega-1} \right) |1,0,0 \rangle ^{\rm{t}} \nonumber \\
& & +\frac{i \gamma_{\rm{t}}}{4} \left(\frac{2\Omega e^{-i t}}{\Omega^2-1}-\frac{e ^{i\Omega t}}{\Omega +1}-\frac{e ^{-i\Omega t}}{\Omega-1} \right) |0,1,0\rangle ^{\rm{t}} \nonumber \\
& & +\frac{i}{8} \left( \frac{2 \Omega e ^{-2 i t}}{\Omega ^2-1}-\frac{e ^{2i\Omega t}}{\Omega+1}-\frac{e ^{-2 i\Omega t}}{\Omega-1} \right) |1,1,0 \rangle ^{\rm{t}} \nonumber \\
& & +\frac{1}{8\sqrt{2}}\left( \frac{2 e ^{-2i t}}{\Omega ^2-1}+\frac{e ^{2i\Omega t}}{\Omega+1}-\frac{e ^{-2 i\Omega t}}{\Omega-1} \right) \left( |2,0,0 \rangle ^{\rm{t}} \right. \nonumber \\
& & \left. \left.-|0,2,0 \rangle ^{\rm{t}} \right) \right],
\label{wvfunc1}
\end{eqnarray}
where $A(t)$ is a constant of normalisation.  From this expression we can deduce that the quadratic average approximation is valid in the linear case, within the parameter regime where $\gamma_{\rm{t}} \ll \Omega$ and $1 \ll \Omega$.

\subsubsection{Nonlinear case}

It is clear that a perturbative two-timescale asymptotic expansion in powers of $1/\Omega$ could be made for the nonlinear case.  Thus, we expect that for the nonlinear case, the quadratic average approximation is also valid within the regime derived above for the linear case.  

We have carried out two dimensional numerial calculations for the nonlinear case which verify this.  For example, for a typical TOP trap system where $r_0=1241$ and $\Omega=153,$ we have propagated the Gross-Pitaevskii equation using both the truncated TOP trap potential of equation (\ref{pot_exp_trunc-3d}) and the potential in the quadratic average approximation, given by equation (\ref{approx-pot-3d}).  An appropriate value for the two dimensional nonlinear stregth is $C_{\rm 2D}=600$ which correspondes to $N\sim2$x$10^4$ in the Otago TOP trap \cite{Wilson03}.  The initial state was chosen to be the ground state of the TOP trap in the quadratic average approximation, i.e.\ the ground state of equation (\ref{Gross-Pitaevskii equation-psi-H}) calculated numerically using optimization methods and shifted in position and momentum according to equations (\ref{meanx_col}) and (\ref{meank_col}) (at $t=0$).  That state was propagated by the Gross-Pitaevskii equation for one period of the bias field rotation for two cases: (i) with the truncated TOP trap potential, giving $\psi^{trunc}(x,y,t=2\pi / \Omega),$ and (ii) with the potential in the quadratic average approximation, yielding $\psi^{quad}(x,y,t=2\pi / \Omega).$  The method used was an accurate fourth order algorithm, with a grid of $512$x$512$ points over a $60$x$60$ range in position, and $20000$ time steps.  The deviation between the two solutions was found to be $\int |\psi^{trunc}(x,y,t=2 \pi /\Omega) - \psi^{quad}(x,y,t=2 \pi /\Omega)  |^2 dxdy=4.44$x$10^{-8}.$

\subsection{Validity of solitary-wave dynamical eigenstates}

The validity of our solitary-wave dynamical eigenstates as dynamical eigenstates of the full TOP trap depends on three validity conditions:
\begin{enumerate}
	\item{the adiabatic approximation,}
    \item{the truncation of the TOP trap potential from equation (\ref{TOP_full}) to equation (\ref{pot_exp_trunc-3d}), and}
    \item{the quadratic average approximation.}
\end{enumerate}
The validity of the quadratic average approximation was addressed above and applies in the regime where $\gamma_{\rm{t}} \ll \Omega$ and $1 \ll \Omega.$  We note that the condition $\gamma_{\rm{t}} \ll \Omega$ can be re-written as $2 r_0 \ll \Omega^3.$  The truncation of the TOP trap potential to yield $V(\mbox{\boldmath$r$},t)$ of equation (\ref{pot_exp_trunc-3d}) is valid provided $x_{\alpha}\ll r_0$ where $x_{\alpha}$ is one of $x$, $y$, or $z$.  A useful estimate of $x_{\alpha}$ is given by the sum of the Thomas-Fermi radius of the solitary-wave dynamical eigenstate envelope wave function with the lowest chemical potential, and the radius of the dynamical eigenstates trajectory, $\gamma_{\rm{t}}$.  This yields
\begin{equation}
x_{\alpha}\approx \left(\frac{30 C}{\sqrt{2} \pi} \right)^{\frac{1}{5}}+\gamma_{\rm{t}},
\end{equation}
where $C$ is given by equation (\ref{Cvalue}).  The adiabatic approximation is valid when the bias field rotation frequency $\Omega$ is much smaller than the Larmor precession frequency, given in our dimensionless units by the potential.  As an estimate of the Larmor precession frequency we use the magnitude of $V_{\rm{TOP}} (\mbox{\boldmath$r$},t)$ which, assuming that $x_{\alpha}\ll r_0,$ is of the order of $r_0^2.$  Thus, the adiabatic approximation is valid provided that $\Omega \ll r_0^2.$  

Finally, collating the validity regimes we find that our solitary-wave dynamical eigenstates calculated using the circularly translating frame, are an accurate description of the dynamical eigenstates of the full TOP trap system within the parameter regime where 
\begin{eqnarray}
2r_0  \ll  & \Omega ^3 & \ll r_0^6 \nonumber\\
1 & \ll & \Omega \label{valid} \\ 
30 C & \ll & \sqrt{2}\pi r_0^5 \nonumber.
\end{eqnarray}
Typical experimental parameters are well within these criteria.  As an example, the Otago TOP trap system of Rubidium-$87$, where $a=55$x$10^{-10}$ m and $\omega_x=18$ Hz \cite{Wilson03}, leads to the third validity condition from equation (\ref{valid}) becoming $N\ll r_0^5$ so all three conditions are easily satisfied.  

\section{The rotating frame \label{sec:rotating}}

In previous theoretical work the {\it rotating frame} has been used to calculate eigenstates of the TOP trap system under various levels of approximation \cite{Kuklov97, Minogin98}.  However, it can be shown that using the rotating frame to describe the TOP trap allows only a limited set of dynamical eigenstates to be found.  For completeness, we present this calculation in appendix~\ref{appendix:rot} where, using the same methods as in sections \ref{sec:circ_trans_frame} and \ref{sec:solitary}, we show that a particular class of solitary-wave solutions in the lab frame are eigenstates of the time independent Gross-Pitaevskii equation in the rotating frame (equation (\ref{Gross-Pitaevskii equation_rot})).  As before, these solitary-wave dynamical eigenstates follow a circular trajectory in the lab frame, described by equations (\ref{meanx_col}) and (\ref{meank_col}), but unlike solitary-wave dynamical eigenstates calculated using the circularly translating frame, the solitary-wave solutions which are eigenstates of equation (\ref{Gross-Pitaevskii equation_rot}) must also obey an additional symmetry, which is that the envelope wave function must be an eigenstate of the $z$ component of angular momentum (see equation (\ref{restriction})).  This condition requires solitary-wave dynamical eigenstates calculated using the rotating frame to be cylindrically symmetric about their centre of mass.  Figure \ref{fig:exstate} shows an example of a dynamical eigenstate envelope wave function with a corresponding solitary-wave dynamical eigenstate which does not satisfy the time independent Gross-Pitaevskii equation in the rotating frame.  Physically we can see why: the nodal line of that solitary-wave dynamical eigenstate, which remains oriented along the $y$ axis in the lab frame, will appear to rotate in the rotating frame so that the dynamical eigenstate is not stationary in that frame.

Solitary-wave dynamical eigenstates of the TOP trap retain their orientation with respect to the lab frame as they move.  Consequently, the rotating frame is not an appropriate choice for describing dynamical eigenstates of the TOP trap system, because the eigenvalue equation in the rotating frame incorporates the angular momentum operator, and places additional symmetry constraints on dynamical eigenstates of the system that are not in general necessary.

\section{Discussion \label{sec:dis}}

We have carried out a detailed characterization of condensate micro-motion in a TOP trap, under some well defined approximations.  These approximations, which are well justified for typical TOP traps, are (i) the adiabatic approximation (which neglects spin precession effects), (ii) the assumption that the condensate is located well within the circle of death, and (iii) the quadratic average approximation (which time-averages quadratically oscillating terms in the potential).  Our treatment allows for condensate nonlinearity and we have shown that within these approximations,  solitary-wave solutions of the nonlinear Gross-Pitaevskii equation exist. We have identified the circularly
translating frame as the most appropriate frame for describing the system, and have shown that eigenstates can be found in that frame, and that they must all be solitary-wave solutions of a certain type.  In particular, all of the solitary-wave dynamical eigenstates have identical centre of mass motion, which in the lab frame is a circular trajectory  with radius $\gamma_{\rm{t}}$ and momentum magnitude $\frac{1}{2}\gamma_{\rm{t}}\Omega$.

Previous theoretical discussions of dynamical eigenstates of the TOP trap have  been given within similar approximations, but with the additional restriction that the nonlinearity due to the atomic interactions is either approximated or neglected.  Kuklov et al.\ \cite{Kuklov97} have obtained exact eigenstates for the linear Schr$\rm{\ddot{o}}$dinger equation within the adiabatic approximation using the truncated TOP trap potential of equation (\ref{pot_exp_trunc-3d}) and those authors have also presented an approximate many-body treatment.  Their exact single particle solutions are obtained using numerous transformations and the form of eigenstate micro-motion is not readily evident.  Their method also employs the rotating frame which, as we have shown within the quadratic average approximation, limits the possible dynamical eigenstates that can be found. Minogin et al.\ \cite{Minogin98} have used an approximate interaction picture method which provides information about the atomic momentum modulation in a TOP trap, but does not describe the micro-motion in the position co-ordinates. 

Our choice of the circularly translating frame allows solitary-wave dynamical eigenstates, which retain their orientation relative to the lab frame, to be readily identified for the TOP trap system.  These dynamical eigenstates have no restriction on the $z$ component of angular momentum of the envelope wave function.  By contrast, we have shown that the dynamical eigenstates calculated using the rotating frame constitute only a subset of the dynamical eigenstates calculated using the circularly translating frame, and are required to be cylindrically symmetric about their centre of mass.

Finally, we have shown that the validity regime for the quadratic average approximation is defined by the conditions $1 \ll \Omega$ and $2r_0 \ll \Omega^3$.  These criteria are well satisfied by existing TOP trap systems.

\acknowledgments{ This work was supported by Marsden Fund 02-PVT-004 and FRST Top Achiever Doctoral Scholarship TAD~884. }

\appendix

\section{Restrictions on solitary-wave solutions which are eigenstates in the circularly translating frame \label{appendix:ctf}}

We seek solutions to equation (\ref{Gross-Pitaevskii equation_tind}) which have the form of equation  (\ref{solution_ctframe}).  We will derive the particular form for the centre of mass motion, $\bar{\mbox{\boldmath$r$}}(t),$ for which such solutions exist.  The centre of mass motion can be solved in general using equation (\ref{Ehrenfest}), which gives
\begin{eqnarray}
\bar{x}(t) = (x_1-\gamma_{\rm{t}})\cos t+v_1 \sin t+\gamma_{\rm{t}} \cos\Omega t  \nonumber \\
\bar{y}(t) = x_2 \cos t+(v_2-\gamma_{\rm{t}}\Omega)\sin t+\gamma_{\rm{t}} \sin\Omega t  \label{centre_mass_motion} \\
\bar{z}(t) = x_3 \cos 2 \sqrt{2} t+\frac{v_3}{2 \sqrt{2}}\sin 2 \sqrt{2}t \nonumber,
\end{eqnarray}
where we have defined $\bar{\mbox{\boldmath$r$}}|_{(t=0)}=(x_1,x_2,x_3)$ and $d\bar{\mbox{\boldmath$r$}}(t)/dt|_{(t=0)}=(v_1,v_2,v_3).$

Eigenstates of the circularly translating frame must satisfy both equation (\ref{Gross-Pitaevskii equation_tind}) and 
\begin{equation}
i \frac{\partial \psi _{\rm{SW}}^{\rm{t}} (\mbox{\boldmath$R$},t)}{\partial t}=\mu_{\rm{SW}}^{\rm{t}} \psi _{\rm{SW}}^{\rm{t}} (\mbox{\boldmath$R$},t).
\label{Gross-Pitaevskii equation_tind_LHS}
\end{equation}
Substituting the solitary wave solution of equation (\ref{solution_ctframe}) into equations (\ref{Gross-Pitaevskii equation_tind}) and (\ref{Gross-Pitaevskii equation_tind_LHS}), and making the change of variables $\mbox{\boldmath$s$}=\mbox{\boldmath$r$}-\bar{\mbox{\boldmath$r$}}(t)$, yields respectively
\begin{eqnarray}
\mu_{\rm{SW}}^{\rm{t}} & = & \mu_{\rm{H}}+\frac{1}{2}\gamma_{\rm{t}}^2\Omega^2+\frac{1}{4}\bar{\mbox{\boldmath$r$}}^2(t)+\frac{7}{4}\bar{z}^2(t)+\frac{1}{4}\left( \frac{d \bar{\mbox{\boldmath$r$}}(t)}{dt} \right)^2 \nonumber \\
& & -\frac{1}{2}\gamma_{\rm{t}} \bar{\mbox{\boldmath$r$}}(t) \cdot (\cos\Omega t,\sin\Omega t,0) +\frac{1}{2} \mbox{\boldmath$s$} \cdot \bar{\mbox{\boldmath$r$}}(t)\nonumber \\
& & +\frac{1}{2}\gamma_{\rm{t}}\Omega \left(\frac{d \bar{\mbox{\boldmath$r$}}(t)}{dt} \right)\cdot (\sin\Omega t,-\cos \Omega t,0) \nonumber \\
& & +\frac{7}{2}\mbox{\boldmath$s$}\cdot (0,0,\bar{z}(t))-\frac{1}{2}\gamma_{\rm{t}} \mbox{\boldmath$s$}\cdot (\cos \Omega t,\sin\Omega t,0) \nonumber \\
& & -i \left[  \frac{d\bar{\mbox{\boldmath$r$}}(t)}{dt} +\gamma_{\rm{t}}\Omega (\sin\Omega t,-\cos\Omega t,0) \right] \cdot \mbox{\boldmath$g$}(\mbox{\boldmath$s$}) \label{RHS_mu},
\end{eqnarray}
and
\begin{eqnarray}
\mu_{\rm{SW}}^{\rm{t}} & = & \mu_{\rm{H}}+\frac{1}{2}\gamma_{\rm{t}}^2\Omega^2-\frac{1}{4}\bar{\mbox{\boldmath$r$}}^2(t)-\frac{1}{2}\bar{\mbox{\boldmath$r$}}(t)\cdot \frac{d^2 \bar{\mbox{\boldmath$r$}}(t)}{dt^2} \nonumber \\
& & +\frac{1}{4}\left( \frac{d \bar{\mbox{\boldmath$r$}}(t)}{dt} \right)^2  -\frac{1}{2}\gamma_{\rm{t}} \Omega^2 \bar{\mbox{\boldmath$r$}}(t) \cdot (\cos\Omega t,\sin\Omega t,0) \nonumber \\ 
& & +\frac{1}{2}\gamma_{\rm{t}}\Omega \left(\frac{d \bar{\mbox{\boldmath$r$}}(t)}{dt} \right)\cdot (\sin\Omega t,-\cos \Omega t,0) \nonumber \\
& &  -\frac{1}{2} \mbox{\boldmath$s$}\cdot \frac{d^2 \bar{\mbox{\boldmath$r$}}(t)}{dt^2}-\frac{1}{2}\gamma_{\rm{t}} \Omega^2 \mbox{\boldmath$s$}\cdot (\cos \Omega t,\sin\Omega t,0) \nonumber \\
& & -i \left[  \frac{d \bar{\mbox{\boldmath$r$}}(t)}{dt} +\gamma_{\rm{t}}\Omega (\sin\Omega t,-\cos\Omega t,0) \right] \cdot \mbox{\boldmath$g$}(\mbox{\boldmath$s$}) \label{LHS_mu},
\end{eqnarray}
where
\begin{equation}
\mbox{\boldmath$g$}(\mbox{\boldmath$s$})=\frac{\nabla_{\mbox{\boldmath$s$}} \psi _{\rm{H}} (\mbox{\boldmath$s$})}{\psi _{\rm{H}} (\mbox{\boldmath$s$})}.
\label{gfuncs}
\end{equation}

Equating equations (\ref{RHS_mu}) and (\ref{LHS_mu}), and substituting the general form of $\bar{\mbox{\boldmath$r$}}(t),$ given by equation  (\ref{centre_mass_motion}), we find that equality requires $x_3=0$ and $v_3=0$ and therefore $\bar{z}(t)=0.$  Inserting this result, and the general expressions for $\bar{x}(t)$ and $\bar{y}(t)$ from equation (\ref{centre_mass_motion}), equations (\ref{RHS_mu}) and (\ref{LHS_mu}) both simplify to 
\begin{eqnarray}
\mu_{\rm{SW}}^{\rm{t}}& = & \mu_{\rm{H}}+\frac{1}{2}\gamma_{\rm{t}}^2\Omega^2 \nonumber \\
& & +\frac{1}{4} (x_1^2+x_2^2+v_1^2+v_2^2-2\gamma_{\rm{t}} (x_1+v_2\Omega)) \nonumber \\
& & +i \mbox{\boldmath$g$}(\mbox{\boldmath$s$}) \cdot ((x_1-\gamma_{\rm{t}})\sin t-v_1 \cos t, \nonumber \\
& & x_2\sin t-(v_2-\gamma_{\rm{t}}\Omega)\cos t,0)   \nonumber \\
& & +\frac{1}{2} \mbox{\boldmath$s$}\cdot ((x_1-\gamma_{\rm{t}})\cos t+v_1 \sin t, \nonumber \\
& & x_2 \cos t+(v_2-\gamma_{\rm{t}}\Omega)\sin t,0) \label{mut_simp}.
\end{eqnarray}
For the solitary-wave solutions of equation (\ref{solution_ctframe}) to be eigenstates of the TOP trap in the circularly translating frame, the chemical potential $\mu_{\rm{SW}}^{\rm{t}}$ must be independent of $\mbox{\boldmath$s$}$ and $t.$  Solving equation (\ref{mut_simp}) for $\mbox{\boldmath$g$}(\mbox{\boldmath$s$}),$ at $t=0$ and $t=\pi/2$, we find that
\begin{eqnarray}
\ln \psi _{\rm{H}}(\mbox{\boldmath$s$}) & = & -\frac{1}{2}i s_1 s_2 \frac{x_2^2+(v_2-\gamma_{\rm{t}}\Omega)^2}{v_1 x_2 -(x_1-\gamma_{\rm{t}})(v_2-\gamma_{\rm{t}}\Omega)}+C_1(s_3) \nonumber \\
 & = &  \frac{1}{2}i s_1 s_2 \frac{(x_1-\gamma_{\rm{t}})^2+v_1^2}{v_1 x_2-(x_1-\gamma_{\rm{t}})(v_2-\gamma_{\rm{t}}\Omega)} \nonumber \\
 & & +C_2(s_3), 
\label{psi_result}
\end{eqnarray}
where $\mbox{\boldmath$s$}=(s_1,s_2,s_3),$ and $C_1(s_3)$ and $C_2(s_3)$ are constants of integration.  The only possible solution is therefore $(x_1,x_2,v_1,v_2)=(\gamma_{\rm{t}},0,0,\gamma_{\rm{t}}\Omega),$ which illiminates $\mbox{\boldmath$g$}(\mbox{\boldmath$s$})$ and $\mbox{\boldmath$s$}$ from equation (\ref{mut_simp}) yielding 
\begin{equation}
\mu_{\rm{SW}}^{\rm{t}}=\mu_{\rm{H}}+\frac{1}{4}\gamma_{\rm{t}}^2(\Omega^2-1),
\end{equation}
which is in agreement with equation (\ref{mu-ct}).  Concluding then, solitary-wave solutions described by equation (\ref{solution_ctframe}) which are eigenstates of the TOP trap in the circularly translating frame exist if and only if $\bar{\mbox{\boldmath$r$}}|_{(t=0)}=(\gamma_{\rm{t}},0,0)$ and $d\bar{\mbox{\boldmath$r$}}(t)/dt|_{(t=0)}=(0,\gamma_{\rm{t}}\Omega,0),$ and therefore the centre of mass motion of these states in the lab frame, given in general by equation (\ref{centre_mass_motion}), simplifies to
\begin{equation}
\label{meanx_col_appendix}
\bar{\mbox{\boldmath$r$}}(t)=\gamma_{\rm{t}} (\cos \Omega t, \sin \Omega t ,0).
\end{equation}

\section{Solitary-wave dynamical eigenstates derived using the rotating frame. \label{appendix:rot}}

In this appendix we calculate dynamical eigenstates of the TOP trap potential in the quadratic average approximation using the rotating frame.  That frame, with co-ordinates $\mbox{\boldmath$r$}'=(x',y',z')$,  rotates at the frequency of the bias field, and is defined by the co-ordinate transformation
\begin{eqnarray}
x' & = & x\cos \Omega t+y\sin \Omega t  \nonumber \\
y' & = & -x\sin \Omega t+y\cos \Omega t \label{rotation} \\
z' & = & z \nonumber.
\end{eqnarray}
In the rotating frame the Gross-Pitaevskii equation becomes
\begin{equation}
\label{Gross-Pitaevskii equation-rot}
i \frac{\partial}{\partial t} \psi ^{\rm{r}} (\mbox{\boldmath$r$}',t)=\mathcal{L}^{\rm{r}}_{\rm{ap}} (\mbox{\boldmath$r$}',t)\psi^{\rm{r}} (\mbox{\boldmath$r$}',t),
\end{equation}
where the evolution operator in the rotating frame is time independent and is given by
\begin{eqnarray}
\mathcal{L}^{\rm{r}}_{\rm{ap}} (\mbox{\boldmath$r$}',t) & = & \mathcal{L}_{\rm{ap}} (\mbox{\boldmath$r$},t)-\Omega \hat{L}_z (\mbox{\boldmath$r$}') \nonumber \\
& = &  -\nabla_{\mbox{\boldmath$r$}'} ^{2}+V^{\rm{r}}_{\rm{ap}}(\mbox{\boldmath$r$}')-\Omega \hat{L}_z (\mbox{\boldmath$r$}') \nonumber \\ & & +C|\psi ^{\rm{r}}(\mbox{\boldmath$r$}',t)|^2,
\label{Lop-rot}
\end{eqnarray}
and the wave function in the rotating frame is $\psi ^{\rm{r}}(\mbox{\boldmath$r$}',t)= \psi (\mbox{\boldmath$r$},t).$  The angular momentum in the rotating frame has a component in the $z$ direction given by
\begin{equation}
\label{ang_mom}
\hat{L}_{z}(\mbox{\boldmath$r$}')=i \left( y'\frac{\partial }{\partial x'}-x'\frac{\partial }{\partial y'}\right),
\end{equation}
and we note that $\hat{L}_{z}(\mbox{\boldmath$r$}')=\hat{L}_{z}(\mbox{\boldmath$r$})$ \cite{L&L}.  Finally, the TOP trap potential of equation (\ref{approx-pot-3d}) becomes, in the rotating frame,
\begin{equation}
\label{pot-rot-pot}
V^{\rm{r}}_{\rm{ap}}(\mbox{\boldmath$r$}')=\frac{1}{4}(x'+2r_0)^2+\frac{1}{4}(y'^2+8z'^2),
\end{equation}
which is a stationary harmonic potential shifted from the origin.

Eigenstates of the TOP trap in the rotating frame obey the time independent Gross-Pitaevskii equation in that frame,
\begin{equation}
\mu^{\rm{r}} \psi ^{\rm{r}} (\mbox{\boldmath$r$}',t)=\mathcal{L}^{\rm{r}}_{\rm{ap}} (\mbox{\boldmath$r$}',t) \psi ^{\rm{r}}  (\mbox{\boldmath$r$}',t).
\label{Gross-Pitaevskii equation_rot}
\end{equation}
Here we show that a particular class of solitary-wave solutions obey equation (\ref{Gross-Pitaevskii equation_rot}).  We denote solitary-wave solutions which are TOP trap eigenstates in the rotating frame by $\psi ^{\rm{r}}_{\rm{SW}} (\mbox{\boldmath$r$}',t),$ with chemical potential $\mu^{\rm{r}}_{\rm{SW}}.$  Transforming the solitary-wave solution of equation (\ref{psi-postulate}) into the rotating frame yields
\begin{eqnarray}
\psi^{\rm{r}}_{\rm{SW}}(\mbox{\boldmath$r$}',t) & = & \psi _{\rm{H}} (\mbox{\boldmath$r$}-\bar{\mbox{\boldmath$r$}}(t)) e ^{-i\mu_{\rm{H}}t+iK(t)} \nonumber \\
& & e^{\frac{1}{2}ix' \left(\cos\Omega t \frac{d\bar{x}(t)}{dt}+\sin\Omega t \frac{d\bar{y}(t)}{dt}\right)} \nonumber \\
& & e^{\frac{1}{2}iy' \left(\cos\Omega t \frac{d\bar{y}(t)}{dt}-\sin\Omega t \frac{d\bar{x}(t)}{dt}\right)} \nonumber \\
& & e^{\frac{1}{2}iz'\left(\frac{d\bar{z}(t)}{dt}\right)}.
\label{sol_wave_rot}
\end{eqnarray}
The detail of substituting equation (\ref{sol_wave_rot}) into equation (\ref{Gross-Pitaevskii equation_rot}) and enforcing $\mu^{\rm{r}}$ to be independent of spatial and temporal co-ordinates is given in section \ref{appendix:rot_rest} of this appendix.  The results are discussed here. 

Solitary-wave dynamical eigenstates of the TOP trap potential, in the quadratic average approximation, as calculated in the rotating frame, have two restrictions.  The first is that the centre of mass motion of the dynamical eigenstates must be given by equations (\ref{meanx_col}) and (\ref{meank_col}).  This is not suprising since we also found this restriction on solitary-wave dynamical eigenstates calculated using the circularly translating frame.  The second restriction is that
\begin{equation}
\hat{L}_z (\mbox{\boldmath$r$}) \psi _{\rm{H}} (\mbox{\boldmath$r$})=l_z \psi _{\rm{H}} (\mbox{\boldmath$r$}),
\label{restriction}
\end{equation}
enforcing the envelope wave function to be an eigenstate of the $z$ component of angular momentum.  This means that dynamical eigenstates, as calculated using the rotating frame, must have a  cylindrically symmetric density about their centre of mass.  We found in our discussion of dynamical eigenstates in section~\ref{sec:eigenstates}, that non-symmetric dynamical eigenstates do exist for the TOP trap and that they are solitary-wave solutions in the lab frame retaining their orientation to that frame.  Solitary-wave dynamical eigenstates which also satisfy equation (\ref{restriction}) are only a subset of the dynamical eigenstates found using the circularly translating frame.

Substituting equations (\ref{meanx_col})and (\ref{restriction}) into the solitary-wave solution in the rotating frame, equation (\ref{sol_wave_rot}), the solitary-wave dynamical eigenstates calculated using the rotating frame, i.e.\ satisfying equation (\ref{Gross-Pitaevskii equation_rot}), have the form
\begin{equation}
\label{psi-rot}
\psi^{\rm{r}}_{\rm{SW}}(\mbox{\boldmath$r$}',t)= \phi(\mbox{\boldmath$r$}') e ^{\frac{1}{2} i \gamma_{\rm{t}} \Omega y'-i \mu^{\rm{r}}_{\rm{SW}} t},
\end{equation}
where we have explicitly extracted the time dependence in the envelope wave function by writing
\begin{equation}
\psi_{\rm{H}}(\mbox{\boldmath$r$}-\bar{\mbox{\boldmath$r$}}(t))=\phi (\mbox{\boldmath$r$}')e ^{i \Omega l_z t}.
\end{equation}
The chemical potential spectrum in the rotating frame is
\begin{equation}
\mu^{\rm{r}}_{\rm{SW}} = \mu_{\rm{H}}+\varepsilon-\Omega l_z,
 \label{mu-rot} 
\end{equation}
where $l_z$ is defined by equation (\ref{restriction}).
The wave function phase $\gamma_{\rm{t}} \Omega y' /2,$ in equation (\ref{psi-rot}), is derived from the co-ordinate dependent phase of equation (\ref{sol_wave_rot}) and accounts for the centre of mass momentum of the eigenstates, given in the lab frame by equation (\ref{meank_col}).  The chemical potential spectrum of the rotating frame, given by equation (\ref{mu-rot}), can be decomposed into three parts: the energy of the state that forms the envelope, the additional energy offset $\varepsilon,$ and an angular momentum term arising from the rotating frame, as expected \cite{L&L}.  
\subsection{Restrictions on solitary-wave solutions which are eigenstates in the rotating frame\label{appendix:rot_rest}}

We seek solutions to equation (\ref{Gross-Pitaevskii equation_rot}) which have the form of equation (\ref{sol_wave_rot}).  The conditions required for such solitary-wave solutions to be eigenstates of the TOP trap in the rotating frame can be found following a similar structure to that used in appendix~\ref{appendix:ctf} for the circularly translating frame.

Solitary-wave solutions which are to be TOP trap eigenstates in the rotating frame must satisfy both equation (\ref{Gross-Pitaevskii equation_rot}) and 
\begin{equation}
i \frac{\partial \psi ^{\rm{r}} _{\rm{SW}} (\mbox{\boldmath$r$}',t)}{\partial t}=\mu^{\rm{r}}_{\rm{SW}} \psi ^{\rm{r}} _{\rm{SW}} (\mbox{\boldmath$r$}',t).
\label{Gross-Pitaevskii equation_tind_LHS_rot}
\end{equation}
Substituting the solitary wave solution of equation (\ref{sol_wave_rot}) into equations (\ref{Gross-Pitaevskii equation_rot}) and (\ref{Gross-Pitaevskii equation_tind_LHS_rot}), and making the change of variables $\mbox{\boldmath$s$}=\mbox{\boldmath$r$}-\bar{\mbox{\boldmath$r$}}(t)$, yields respectively
\begin{eqnarray}
\mu^{\rm{r}}_{\rm{SW}} & = & 
\mu_{\rm{H}}-\frac{1}{2}\bar{\mbox{\boldmath$r$}}(t) \cdot  \frac{d^2 \bar{\mbox{\boldmath$r$}}(t)}{dt^2} +\frac{1}{4} \bar{\mbox{\boldmath$r$}}^2 (t)+\frac{7}{4}\bar{z}^2(t) \nonumber \\
& & +\bar{\mbox{\boldmath$r$}}(t)\cdot r_0(\cos\Omega t,\sin\Omega t,0)+\frac{7}{2}\mbox{\boldmath$s$}\cdot (0,0,\bar{z}(t)) \nonumber \\
& & 
+\frac{1}{2} \mbox{\boldmath$s$} \cdot \left[\bar{\mbox{\boldmath$r$}}(t)+2r_0(\cos\Omega t,\sin\Omega t,0)\right] \nonumber \\
& & +\frac{1}{2}\Omega \left( \frac{d \bar{\mbox{\boldmath$r$}}(t)}{dt} \right) \cdot (\bar{y}(t),-\bar{x}(t),0) +\frac{1}{4} \left( \frac{d \bar{\mbox{\boldmath$r$}}(t)}{dt} \right)^2 \nonumber \\
& & -\frac{1}{2} \Omega \mbox{\boldmath$s$}  \left( \frac{d \bar{y}(t)}{dt},-\frac{d \bar{x}(t)}{dt},0 \right) -\Omega l_z(\mbox{\boldmath$s$}) \nonumber \\
& & -i \left[  \frac{d\bar{\mbox{\boldmath$r$}}(t)}{dt} +\Omega (\bar{y}(t),-\bar{x}(t),0) \right] \cdot \mbox{\boldmath$g$}(\mbox{\boldmath$s$}) \label{RHS_mu_rot},
\end{eqnarray}
and
\begin{eqnarray}
\mu^{\rm{r}}_{\rm{SW}} & = &
\mu_{\rm{H}}-\frac{1}{2} \bar{\mbox{\boldmath$r$}}(t) \cdot  \frac{d^2 \bar{\mbox{\boldmath$r$}}(t)}{dt^2} -\frac{1}{4}\bar{\mbox{\boldmath$r$}}^2 (t)-\frac{1}{2} \mbox{\boldmath$s$} \cdot \frac{d^2 \bar{\mbox{\boldmath$r$}}(t)}{dt^2} \nonumber \\
& & +\frac{1}{2}\Omega \left( \frac{d \bar{\mbox{\boldmath$r$}}(t)}{dt} \right) \cdot (\bar{y}(t),-\bar{x}(t),0)+\frac{1}{4} \left( \frac{d \bar{\mbox{\boldmath$r$}}(t)}{dt} \right)^2 \nonumber \\
& & -\frac{1}{2}\Omega \mbox{\boldmath$s$} \cdot \left( \frac{d \bar{y}(t)}{dt},-\frac{d \bar{x}(t)}{dt},0 \right) -\Omega l_z(\mbox{\boldmath$s$})\nonumber \\
& & -i \left[  \frac{d\bar{\mbox{\boldmath$r$}}(t)}{dt} +\Omega (\bar{y}(t),-\bar{x}(t),0) \right] \cdot \mbox{\boldmath$g$}(\mbox{\boldmath$s$})
\label{LHS_mu_rot},
\end{eqnarray}
where $\mbox{\boldmath$g$}(\mbox{\boldmath$s$})$ is given by equation (\ref{gfuncs}) and 
\begin{equation}
l_z(\mbox{\boldmath$s$})=\frac{\hat{L}_{z} (\mbox{\boldmath$s$}) \psi _{\rm{H}} (\mbox{\boldmath$s$})}{\psi _{\rm{H}} (\mbox{\boldmath$s$})}.
\end{equation}

Equating equations (\ref{RHS_mu_rot}) and (\ref{LHS_mu_rot}), and substituting the general form of $\bar{\mbox{\boldmath$r$}}(t),$ given by equation  (\ref{centre_mass_motion}), we find that equality requires $x_3=0$ and $v_3=0$ and therefore $\bar{z}(t)=0.$  Inserting this result, and the general expressions for $\bar{x}(t)$ and $\bar{y}(t)$ from equation (\ref{centre_mass_motion}), equations (\ref{RHS_mu_rot}) and (\ref{LHS_mu_rot})  both simplify to 
\begin{eqnarray}
\mu^{\rm{r}}_{\rm{SW}} & = & 
\mu_{\rm{H}}-\Omega l_z(\mbox{\boldmath$s$})+\frac{1}{4} (x_1^2+x_2^2+v_1^2+v_2^2) \nonumber \\
& & +\frac{1}{2}( \Omega (v_1x_2-v_2x_1)+\gamma_{\rm{t}} x_1 (\Omega^2-1)) \nonumber \\
& & -i \mbox{\boldmath$g$}(\mbox{\boldmath$s$}) \cdot (c_1\cos t+c_2\sin t,c_3\cos t-c_4\sin t,0)   \nonumber \\
 & & -\frac{1}{2} \mbox{\boldmath$s$}\cdot (c_2\cos t-c_1 \sin t, \nonumber \\
 & & -c_4 \cos t-c_3\sin t,0),
\label{mut_simp_rot}
\end{eqnarray}
where $c_1=v_1+\Omega x_2,$ $c_2=v_2\Omega-x_1-\gamma_{\rm{t}}(\Omega^2-1),$ $c_3=v_2-\Omega x_1,$ and $c_4=v_1\Omega+x_2.$  For the solitary-wave solutions in the lab frame to be eigenstates of the TOP trap in the rotating frame, the chemical potential $\mu^{\rm{r}}_{\rm{SW}}$ must be independent of $\mbox{\boldmath$s$}$ and $t.$  Solving for $\mbox{\boldmath$g$}(\mbox{\boldmath$s$}),$ at $t=0$ and $t=\pi/2$, we find that the only possible solution occurs when $c_1=c_2$ and $c_3=-c_4.$  Substituting back into equation (\ref{mut_simp_rot}) we find that $(c_1,c_2,c_3,c_4)=(0,0,0,0),$ or rather $(x_1,x_2,v_1,v_2)=(\gamma_{\rm{t}},0,0,\gamma_{\rm{t}}\Omega).$  Furthermore, we find that $l_z(\mbox{\boldmath$s$})$ must be independent of $\mbox{\boldmath$s$},$ and the chemical potential in the rotating frame is 
\begin{equation}
\mu^{\rm{r}}_{\rm{SW}}=\mu_{\rm{H}}+\frac{1}{4}\gamma_{\rm{t}}^2(\Omega^2-1)-\Omega l_z,
\end{equation}
which is in agreement with equation (\ref{mu-rot}).  Concluding then, solitary-wave solutions which are eigenstates of the TOP trap potential in the rotating frame exist if and only if $\bar{\mbox{\boldmath$r$}}|_{(t=0)}=(\gamma_{\rm{t}},0,0),$  $d\bar{\mbox{\boldmath$r$}}(t)/dt|_{(t=0)}=(0,\gamma_{\rm{t}}\Omega,0),$ and 
\begin{equation}
\hat{L}_z (\mbox{\boldmath$s$}) \psi _{\rm{H}} (\mbox{\boldmath$s$})=l_z \psi _{\rm{H}} (\mbox{\boldmath$s$}).
\label{restriction_app}
\end{equation}
The centre of mass motion in the lab frame of solitary-wave dynamical eigenstates calculated using the rotating frame is identical to that of solitary-wave dynamcial eigenstates found using the circularly translating frame.  Equation (\ref{sol_wave_rot}) can now be written as
\begin{equation}
\psi ^{\rm{r}}_{\rm{SW}}(\mbox{\boldmath$r$}',t)=\psi_{\rm{H}}(\mbox{\boldmath$r$}-\bar{\mbox{\boldmath$r$}}(t)) e^{\frac{1}{2}i \gamma_{\rm{t}}\Omega y'-i (\mu_{\rm{H}}+\varepsilon) t },
\end{equation}
and using equations (\ref{Gross-Pitaevskii equation_tind_LHS_rot}) and (\ref{restriction_app}) it can be shown that
\begin{equation}
\psi_{\rm{H}}(\mbox{\boldmath$r$}-\bar{\mbox{\boldmath$r$}}(t))=\phi (\mbox{\boldmath$r$}')e ^{i \Omega l_z t}.
\end{equation}


\begin{thebibliography}{15}
\bibitem{Cornell95} Wolfgang Petrich, Michael H Anderson, Jason R Ensher and Eric A Cornell {\it Phys. Rev. Lett.} {\bf 74} 3352 (1995)
\bibitem{Arimondo00}  J H M$\rm{\ddot{u}}$ller, O Morsch, D Ciampini, M Anderlini, R Mannella and E Arimondo {\it Phys. Rev. Lett.} {\bf 85} 4454 (2000)
\bibitem{Arimondo01} J H M$\rm{\ddot{u}}$ller, O Morsch, D Ciampini, M Anderlini, R Mannella and E Arimondo {\it C. R. Acad. Sci. Paris} {\bf 2} 649 (2001)
\bibitem{Kuklov97} A B Kuklov, N Chencinski, A M Levine, W M Schreiber and Joseph L Birman {\it Phys. Rev. A} {\bf 55} 488 (1997) 
\bibitem{Minogin98} Vladimir G Minogin, James A Richmond and Geoffrey I Opat {\it Phys. Rev. A} {\bf 58} 3138 (1998) 
\bibitem{Shtrikman99} S Gov and S Shtrikman {\it J. Appl. Phys.} {\bf 86} 2250 (1999)
\bibitem{Arimondo04} Roberto Franzosi, Andrea Spinelli, Bruno Zambon and Ennio Arimondo {\it cond-mat/0403466} (2004)
\bibitem{Arimondo02} M Cristiani, O Morsch, J H M$\rm{\ddot{u}}$ller, D Ciampini and E Arimondo {\it Phys. Rev. A} {\bf 65} 063612 (2002)
\bibitem{Wilson03}  R Geursen, N R Thomas and A C Wilson {\it Phys. Rev. A}
 {\bf 68} 043611 (2003) 
\bibitem{Metcalf85} Alan L Migdall, John V Prodan, William D Phillips, Thomas H Bergeman and Harold J Metcalf {\it Phys. Rev. Lett.} {\bf 54} 2596 (1985)
\bibitem{Metcalf87} T Bergeman, Gidon Erez and Harold J Metcalf {\it Phys. Rev. A} {\bf 35} 1535 (1987)
\bibitem{Morgan97} S A Morgan, R J Ballagh and K Burnett {\it Phys. Rev. A} {\bf 55} 4338 (1997)
\bibitem{Margetis99} Dionisios Margetis {\it J. Math. Phys.} {\bf 40} 5522 (1999)
\bibitem{Band02} Y Japha and Y B Band {\it J. Phys. B: At. Mol. Opt. Phys.} {\bf 35} 2383 (2002)
\bibitem{L&L} L D Landau and E M Lifshitz {\it Course of  Theoretical Physics Volume 1 Mechanics} (Oxford: Pergamon Press Ltd) p~126 (1960)
\end{thebibliography}
\end{document}